\documentclass[aps,prl,reprint,superscriptaddress,twocolumn,showpacs]{revtex4-1}
\pdfoutput=1
\usepackage{amsfonts}
\usepackage{amsmath}
\usepackage{multirow}
\usepackage{longtable}
\usepackage{graphicx}

\newcommand{\be}{\begin{equation}}
\newcommand{\ee}{\end{equation}}

\newcommand{\av}[1]{\langle #1 \rangle}

\begin{document}



\title{The Scaling of Human Contacts in Reaction-Diffusion Processes on Heterogeneous Metapopulation Networks}


\author{Michele Tizzoni}
\email[]{michele.tizzoni@isi.it}
\affiliation{Computational Epidemiology Laboratory, ISI Foundation, Torino, via Alassio 11/C, Italy}

\author{Kaiyuan Sun}
\email[]{k.sun@neu.edu}
\affiliation{Laboratory for the Modeling of Biological and Socio-technical Systems,
Northeastern University, Boston MA 02115 USA}

\author{Diego Benusiglio}
\email[]{diego.benusiglio@gmail.com}
\affiliation{Computational Epidemiology Laboratory, ISI Foundation, Torino, via Alassio 11/C, Italy}
\affiliation{Dipartimento di Fisica, Universit\'a degli Studi di Torino, via Giuria 1, Torino, Italy}

\author{M\'arton Karsai}
\email[]{marton.karsai@ens-lyon.fr}
\affiliation{Laboratoire de l'Informatique du Parall\'elisme, INRIA-UMR 5668, IXXI, ENS de Lyon, 69364 Lyon, France}

\author{Nicola Perra}
\email[]{n.perra@neu.edu}
\affiliation{Laboratory for the Modeling of Biological and Socio-technical Systems,
Northeastern University, Boston MA 02115 USA}


\date{\today}

\begin{abstract}
We present new empirical evidence, based on millions of interactions on Twitter, confirming that human contacts scale with population sizes. We integrate such observations into a reaction-diffusion metapopulation framework providing an analytical expression for the global invasion threshold of a contagion process. Remarkably, the scaling of human contacts is found to facilitate the spreading dynamics. Our results show that the scaling properties of human interactions can significantly affect dynamical processes mediated by human contacts such as the spread of diseases, and ideas.
\end{abstract}


\pacs{$89.75.k$, $05.70.Ln$, $87.23.Ge$}

\maketitle

In the past fifteen years network theory has developed a wide range of mathematical tools to study and model dynamical processes on complex networks \cite{Barrat2008, Newman2010, Havlin2010}.
In particular, building upon a long research tradition in ecology \cite{Hanski1997}, the theoretical framework of reaction-diffusion (RD) processes on metapopulation networks has been proved to be extremely valuable for describing contagion processes in spatially structured systems \cite{Pastorsatorras2014}. In the RD metapopulation framework, individuals are represented by particles that reside in nodes of a network and can migrate along the connections between them. Each node describes a subpopulation, i.e. a city or a town, while each link represents a travel route. Inside each node, particles react according to the rules of the process under study. 
Such modeling framework has been widely used to describe the dynamics of a number of real world complex systems \cite{Gallos2004, Holland2008, Nakao2010}.
Its most successful application, though, has been the modeling of the spread of infectious diseases in structured populations \cite{Colizza2007b, Colizza2007c, Colizza2008, Liu2013,Hufnagel2004, Balcan2009b, Balcan2009a, Ferguson2005,Merler2011, Tizzoni2012}. 
A common assumption in RD metapopulation models is that particles interact in each node with the same contact rate, constant and equal for any given size of the subpopulation. In mathematical epidemiology, such assumption corresponds to the frequency-dependent transmission rate \cite{Hu2013}. 
However, a recent study based on the analysis of large mobile phone datasets \cite{Schlapfer2014} has showed evidence that the {\it per capita} social connectivity scales with the subpopulation size. 
In particular, the authors of \cite{Schlapfer2014} found that the cumulative number of social contacts $K$ of individuals in a city scales as $K \sim N^{\gamma}$ where $\gamma>1 $ and $N$ is the city's population. 
This finding is consistent with a number of scaling properties observed in cities \cite{Bettencourt2007, Calabrese2011c} and with theoretical models of urban development \cite{Pan2013}.\\
In this Letter, we first present new empirical evidence, based on the analysis of human interactions on Twitter that supports the contacts scaling hypothesis. 
Then, we integrate such observation into a RD metapopulation framework characterized by realistic heterogeneities in the distribution of the number of connections per node and in traffic flows. In particular, we study a Susceptible-Infectious-Recovered (SIR) epidemic dynamics inside each subpopulation~\cite{Anderson1992}. 
We provide an explicit analytical expression for the global invasion threshold that sets a critical value of the diffusion/mobility rate below which a contagion process is not able to spread to a macroscopic fraction of the system~\cite{Colizza2008}. We show that the scaling of interaction rates with subpopulation size significantly alters the contagion dynamics leading to a lower critical value of the mobility rate. Interestingly, such variations are enhanced by increasing heterogeneities in mobility patterns coupling the subpopulations. Given the applicability of the RD metapopulation framework to a wide range of phenomena such as knowledge diffusion, opinion and infectious disease spread, our results open the way to the inclusion of more realistic interaction patterns in the modeling of such contagion processes.

\begin{figure}[!t]
\includegraphics[width=0.4\textwidth]{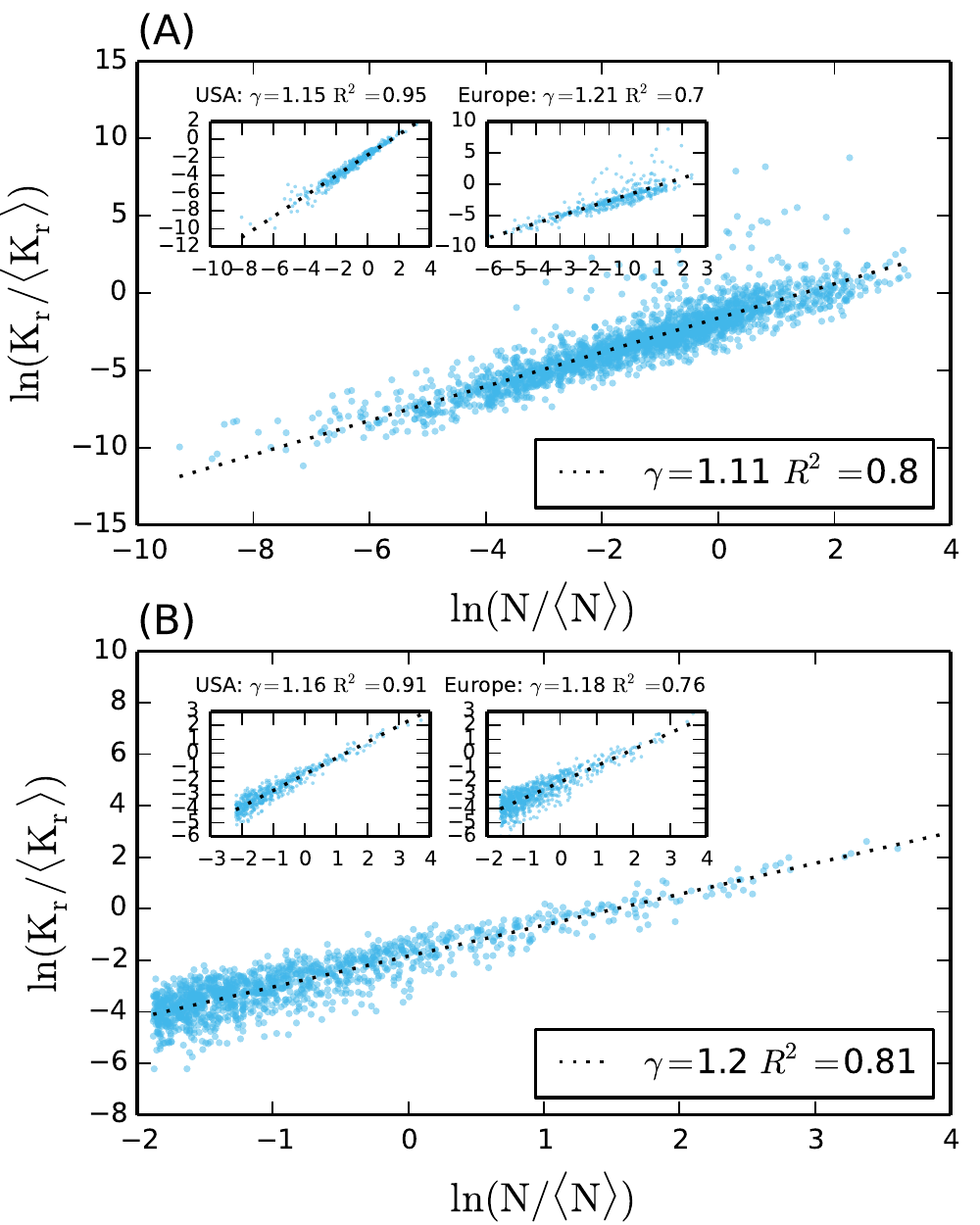}
\caption{\label{fig:data} (A) Rescaled cumulative degree $K_r$ against population $N$, measured between $13\,129\,406$ Twitter users distributed across $2371$ basins in $205$ countries. (B) Rescaled cumulative degree against population, measured between $4\,606\,444$ Twitter users in $1344$ metropolitan areas in $31$ countries. We normalized the values of $K_r$ and $N$ by their average to compare the results across different countries. Insets show the dependency of $K_r$ on $N$ restricted to the Twitter users in the US and Europe.}
\end{figure}

We analyze the interactions between users of the micro-blogging platform Twitter in several countries. The empirical measurements of scaling behavior at the population level are known to be affected by the definition of the boundaries of the census areas~\cite{Louf2014a}. 
For this reason we considered two different geographical aggregations. The first maps about $13$ millions Twitter users into $2371$ census areas centered around major transportation hubs~\cite{Balcan2009b} in $205$ countries. 
Such aggregation level has been used to model the spreading of infectious diseases at the global scale~\cite{Balcan2009a,Tizzoni2012}. The second maps about $4.6$ million Twitter users into $1344$ metropolitan areas, across the USA and $31$ European countries. 
See the Supplemental Material for further details about the data and the geographic aggregations. To extract the relation between contacts and population size, we follow the methods used by Schl\"apfer and colleagues in their analysis \cite{Schlapfer2014}.
In particular, in both aggregation levels we build the reciprocal communication network through Twitter mention interactions: a link is placed between users $A$ and $B$ within a given census area if and only if $A$ mentioned $B$ and $B$ mentioned $A$ back at least once. Similar results are obtained considering also the connections outside the census area (see the Supplemental Material for details). 
We calculate the total number of links $K=\sum_{i\in S} k_i$, where $S$ is the number of users within a census area, and rescale it by the users' coverage $\frac{S}{N}$ to obtain $K_r=K\frac{N}{S}$. In both cases, we find, consistently with Schl\"apfer \cite{Schlapfer2014}, that the rescaled cumulative degree $K_r$ is characterized by a power-law relation with the population of the census areas, $K_r \propto N^{\gamma}$ with exponent $\gamma=1.11 \pm 0.01$ considering basins and $\gamma=1.20 \pm 0.02$ considering metropolitan areas (see Fig. \ref{fig:data}).
We also restrict our analysis of the Twitter dataset to the two aggregation levels in the USA and Europe. We find that the scaling behavior still holds, with the exponent $\gamma$ in the same range, i.e. $\gamma=1.15 \pm 0.01$ in the USA and $\gamma=1.21 \pm 0.04$ in Europe considering census areas, and $\gamma=1.16 \pm 0.02$ in the USA and $\gamma=1.18 \pm 0.02$ in Europe considering metropolitan areas. In the Supplemental Material, we report all the details of the data analysis.

To study the effect of the scaling of contact rates in RD processes, we consider a metapopulation network of $V$ nodes, where each node $i$ is characterized by a subpopulation of $N_i$ individuals and degree $k_i$, representing the number of subpopulations connected to it. 
We adopt a degree-block approximation, assuming all the subpopulations of degree $k$ to be statistically equivalent \cite{Colizza2007b, Colizza2008,Colizza2007c} and we denote the degree distribution of the network as $P(k)$.
To describe the diffusion of individuals, we assume that the rate at which individuals leave a subpopulation is independent of its degree and equal to $p$. However, to reproduce the properties of real networks~\cite{Barrat2004}, we consider a heterogeneous distribution of traffic flows along any given connection. In particular, the diffusion rate of individuals between two nodes of degree $k$ and $k'$ is $d_{kk'}=p \frac{w_0(k'k)^{\theta}}{T_{k}}$, where $T_{k}$ provides the necessary normalization. It is possible to show that, under such conditions, the population size of a node of degree $k$, $N_k$, at equilibrium is given by $N_{k}=\bar{N}\frac{k^{1+\theta}}{\av{k^{1+\theta}}}$, where $\bar{N}=\sum_kP(k)N_{k}(t)$~\cite{Colizza2008}. As a consequence, the exponent $\theta$, which modulates the heterogeneity of the mobility flows, also regulates the heterogeneity of the subpopulations size distribution.
We model the reactions, taking place in each node, as a stochastic SIR epidemic process defined by a transmissibility $\lambda$ and recovery rate $\mu$~\cite{Keeling2008}. In each subpopulation individuals are partitioned in three compartments according to their health status: susceptibles (S), infectious (I) and recovered (R). 
The SIR dynamics are defined by two transitions~\cite{Keeling2008}. The first describes the infection process: $S+I \rightarrow 2I$, while the second describes the recovery process: $I \rightarrow S$. Here, we investigate the case in which the infection dynamics is dependent on the local population size. 
More precisely, inside each node, we consider an homogeneous mixing approximation where the average contact rate scales with the population size as $\langle k \rangle \sim N^{\eta}$.
The values of the exponent $\gamma$ measured in real social networks correspond to $\eta=\gamma-1$ ranging between $0.11$ and $0.2$. The value of $\eta$ measured in \cite{Schlapfer2014} is $\eta=0.12$. In the following, without lack of generality, we focus on the case $\eta>0$.  
The immediate consequence of this assumption is that the basic reproductive number $R_0$, i.e. the average number of newly infected individuals generated by an infectious one in a fully susceptible population~\cite{Anderson1992}, depends on the population size as:
\be
\label{eq:R0_fun}
R_0(k)=\frac{\lambda}{\mu}N^{\eta}_k=\frac{\lambda}{\mu}\bar{N}^{\eta}\frac{k^{(1+\theta)\eta}}{\av{k^{1+\theta}}^{\eta}}=\mathcal{M}k^\xi \,,
\ee
where $\mathcal{M} = \frac{\lambda}{\mu}\frac{\bar{N}^{\eta}}{\av{k^{1+\theta}}^{\eta}}$ is a constant that depends on the characteristics of the disease and the metapopulation structure (see the Supplemental Material for the complete derivation).
It is immediate to see from Eq. \ref{eq:R0_fun} that the reproductive number will significantly vary from  one location to another, depending on the degree of each node and on the exponent $\xi=(1+\theta)\eta$, which combines the heterogeneity of the traffic flows and of the contact rates.
\begin{figure}[!t]
\includegraphics[width=0.4\textwidth]{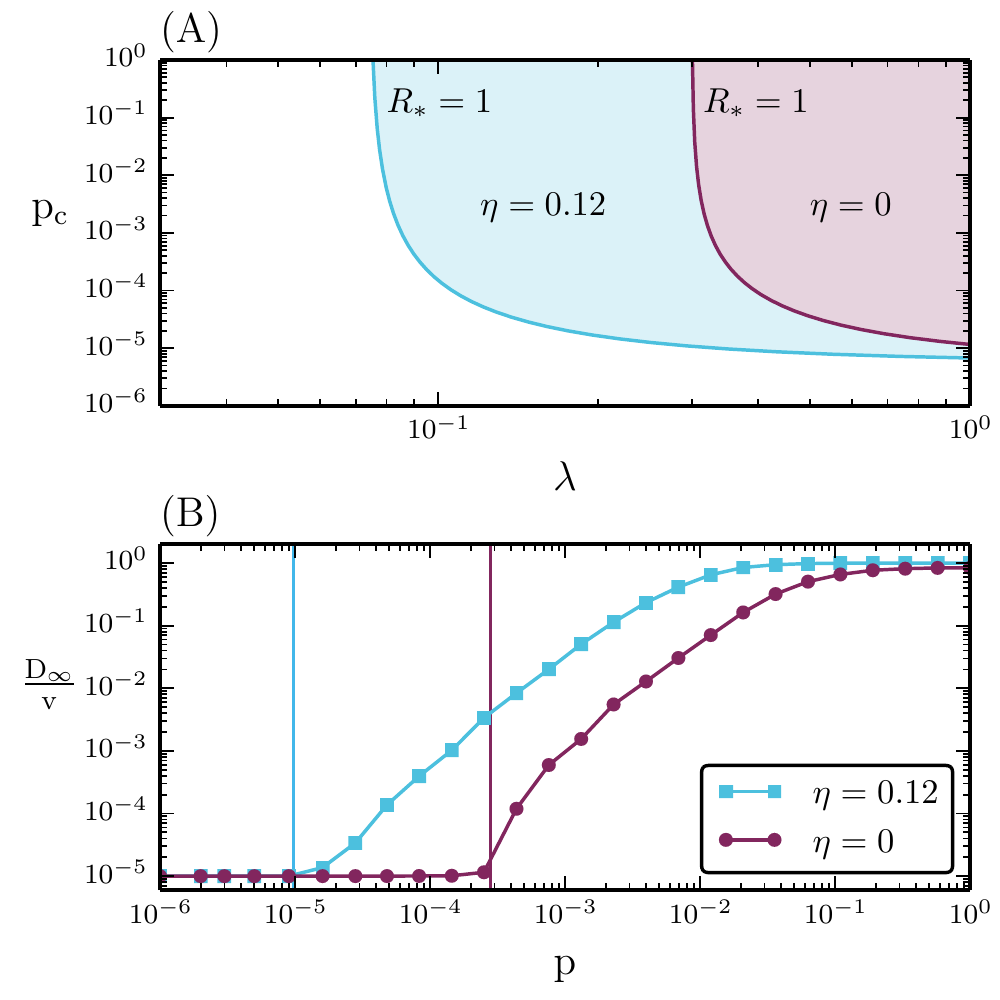}
\caption{\label{fig:eta} (A) Phase diagram defined by the threshold condition $R_*(p,\lambda)=1$ for $\eta=0$ and $\eta=0.12$. We consider uncorrelated scale-free networks of $V=10^5$ nodes, and $P(k) \sim k^{-2.1}$. We set $\theta=0.5$, $\bar{N}=10^3$, and $\mu=0.3$. (B) Simulated global attack rate $D_{\infty}/V$ as a function of the mobility rate $p$ for different values of the contact scaling exponent $\eta$ and $\lambda=0.35$. Vertical lines indicate the critical threshold value $p_c$ calculated by setting $R_*=1$ Eq. \ref{eq:globalth}. Each point is averaged over $2 \times 10^3$ simulations.}
\end{figure} 
The necessary and sufficient condition for the local spreading of the disease in nodes of degree $k$ is given by the local epidemic threshold, i.e. $R_0(k)>1$. It is important to notice that this may not be satisfied in all the subpopulations. Such situation is realistic for a number of epidemic scenarios where, due to specific characteristics of the local population, the value of the basic reproductive varies across locations~\cite{Dalziel2014}. 
The crucial question in metapopulations systems is evaluating the necessary conditions under which a local epidemic outbreak leads to a global outbreak. 
This implies defining an invasion threshold $R_*$ for the whole system~\cite{Colizza2008}. In order to find an analytical expression for $R_*$, we describe the epidemic invasion as a branching process~\cite{Ball1997, Colizza2007c, Colizza2008, Balcan2011, Poletto2012, Bajardi2011a} relating the number of subpopulations of degree $k$ that have been reached by the epidemic at generation $n$, $D_k^n$, with $D_k^{n-1}$:
\be
\label{eq:general}
\small
D_k^n=\sum_{k'}D_{k'}^{n-1}(k'-1)P(k|k')\left ( 1-R_0(k)^{-\lambda_{k'k}} \right) \left ( 1- \frac{D_k^{n-1}}{V_k}\right ).
\ee
The term $k'-1$ considers that each diseased subpopulation of degree $k'$ and generation $n-1$, $D_{k'}^{n-1}$, can seed all the connected nodes but the one from which it received the infection. The term $P(k|k')$ describes the probability that nodes of degree $k'$ are connected with nodes of degree $k$. We consider uncorrelated networks where this conditional probability does not depend on $k'$ and $P(k|k')=kP(k)/\langle k \rangle$. 
The term $1-R_0(k)^{-\lambda_{k'k}}$ defines the probability that, given $\lambda_{k'k}$ infectious individuals seeding a node of degree $k$, the subpopulation will experience a local outbreak~\cite{Bailey1975}. This number can be estimated as:
\be
\lambda_{k'k}=2\frac{R_0(k')-1}{R_0(k')^2}N_{k'}\times \frac{1}{\mu} \times d_{k'k} \times \delta[R_0(k')].
\ee
Indeed, the total number of infected individuals generated at the source can be approximated as $2\frac{R_0(k')-1}{R_0(k')^2}N_{k'}$~\cite{Colizza2008}, infectious individuals recover, on average, after $\mu^{-1}$ time steps, and the diffusion rate between the two degree classes is $d_{k'k}$. It is important to notice that such approximations are valid only for $R_0(k')>1$. 
Indeed, if this condition is not satisfied the disease will not be able to spread locally in any subpopulation $k'$. To address this issue, we introduce the a step function:
\be
\delta[R_0(k')] = \left\{ 
  \begin{array}{l l}
    1 & \quad \text{for} \; k'\;|\; R_0(k')>1 \\
    0 & \quad \text{for} \; k'\;|\; R_0(k')<1
  \end{array} \right.
\ee
Finally, the last term in Eq.~\ref{eq:general} represents the fraction of subpopulations of degree $k$ that are not yet infected.
By plugging all these terms in Eq. \ref{eq:general}, it is possible to solve it analytically (see details in the Supplemental Material) and find an explicit expression for the global epidemic threshold:
\begin{eqnarray}
R_*&=&\frac{2p\bar{N}}{\mu}\frac{1}{\av{k^{1+\theta}}^2} [ \av{k^{2+2\theta}}^* - \av{k^{1+2\theta}}^*\\ \nonumber
&-& \frac{2}{\mathcal{M}}[ \av{k^{2+2\theta-\xi}}^*-\av{k^{1+2\theta-\xi}}^*]\\ \nonumber
&+& \frac{1}{\mathcal{M}^2} [ \av{k^{2+2\theta-2\xi}}^*-\av{k^{1+2\theta-2\xi}}^*]\\ \nonumber
&\equiv&\frac{2p\bar{N}}{\mu}\mathcal{F}(P(k),\theta,\eta,\lambda,\mu,\bar{N}).
\label{eq:globalth}
\end{eqnarray}
All the moments denoted by a star are calculated over a subset of degree values. More specifically, we define the general starred degree moment as $\av{k^{x}}^*=\sum_{k} \delta[R_0(k)] k^{x}P(k)$. The function $\mathcal{F}$ describes the dependence of the threshold on the properties of the network, the mobility patterns, the scaling of contacts, and the details of the disease.
Interestingly, the denominator factor $\av{k^{1+\theta}}^2$ is related to the mobility between subpopulations and not to the spreading dynamics within nodes, therefore the corresponding moment of the degree distribution is calculated over all the values of $k$. 

The expression of the global invasion threshold defines the range of parameters for which a global outbreak is possible, corresponding to the solutions of the equation $R_*=1$. 
For $R_*<1$ an outbreak seeded in any subpopulation will eventually die, while for $R_*>1$ the contagion process will eventually reach a finite fraction of the system with non-zero probability.
Since our focus is the interplay between the heterogeneity of contact rates and the mobility rates, we look at the effect of the parameter $\eta$ compared to the case $\eta=0$ that has been previously studied \cite{Colizza2007c, Colizza2008}.
Indeed, from Eq. \ref{eq:globalth} it possible to see that, by setting $\eta=\xi=0$, we consistently recover the same expression of $R_*$ derived in the case of a constant contact rate across subpopulations \cite{Colizza2008}.
In particular, we compare the value of the critical mobility rate $p_c$, corresponding to the solution of $R_*=1$ ($p_c=\frac{\mu}{2\bar{N}} \mathcal{F}^{-1}$), in the two cases: $\eta>0$ and $\eta=0$.
The introduction of a scaling contact rate in every subpopulation, modifies the result of Ref. \cite{Colizza2008} by increasing the overall heterogeneity of the metapopulation system and, eventually, by reducing the critical value of $p$.
More specifically, values of $\eta>0$ as observed from empirical social networks, alter the spreading dynamics by accelerating the contagion process and thus increasing the value of $R_*$. 
This implies that, for a given set of parameters describing the mobility network, the metapopulation system and the transmissibility of the infectious agent, the critical mobility value will be lower for larger values of $\eta$.
Fig. \ref{fig:eta}a shows the invasion region in the plane $R_*(p,\lambda)$ for $\eta=0$ and $\eta=0.12$, with the latter clearly displaying a larger portion of the phase space in the global spreading regime. In particular, the scaling of contacts with subpopulation sizes allows the global spreading of diseases characterized by significantly smaller values of transmissibility $\lambda$.
We confirm our analytical findings through extensive numerical simulations performed considering uncorrelated scale-free networks with $V=10^5$ nodes, and exponent $\gamma=-2.1$~\cite{Catanzaro2005}. In Fig. \ref{fig:eta}b, we compare the global attack rate, i.e. the final fraction of subpopulations that experienced a local outbreak, for two identical metapopulation structures and different values of $\eta$. The results of $2 \times10^3$ Monte Carlo simulations per point show an excellent agreement with the theoretical threshold calculated from 
Eq. \ref{eq:globalth}.

\begin{figure}[!t]
\includegraphics[width=0.45\textwidth]{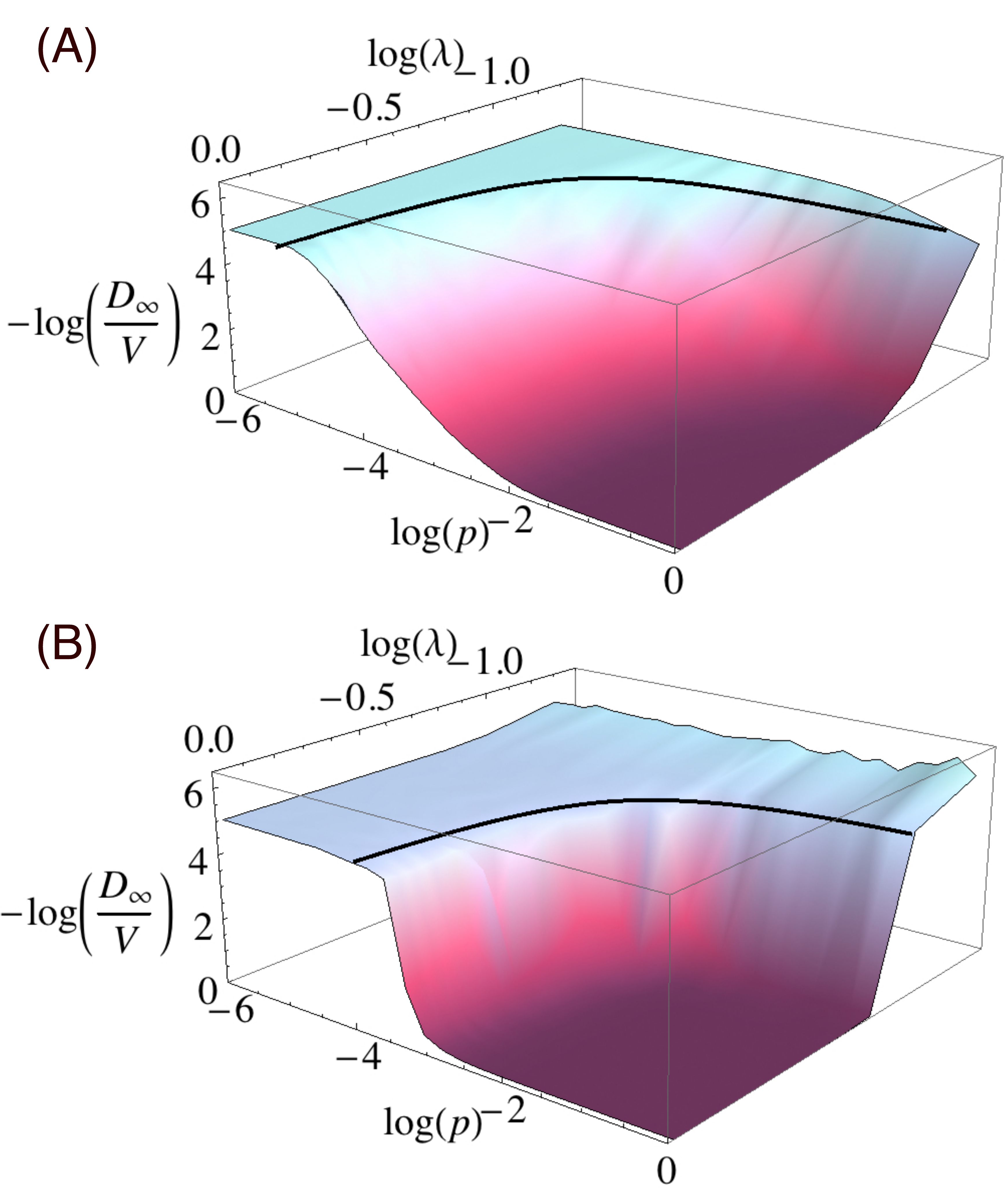}
\caption{\label{fig:phase_space} Simulated global attack rate $D_{\infty}/V$ as a two-dimensional function of the mobility rate $p$ and the transmissibility $\lambda$ for different mobility network structures characterized by $\theta=0.5$ (A) and $\theta=-0.4$ (B). Black solid lines indicate the analytical predictions for the critical values of $p$ and $\lambda$ corresponding to $R_*=1$. Here the network parameters are the same of Fig. \ref{fig:eta} and $\eta=0.12$. Each point of the phase-space is averaged over $2000$ realizations of the model.  To facilitate the visual comparison between the simulations and the analytical solutions we plot the $z$-axis considering the negative $\log_{10}$ of $D_\infty/V$.}
\end{figure}

Overall, the global epidemic threshold is determined in a non-linear way, through the exponent $\xi$, by the interplay between the contact rate heterogeneity, tuned by the exponent $\eta$, and the heterogeneity of the mobility patterns, tuned by the exponent $\theta$. 
The latter can be tuned to counterbalance the effect of the contact scaling on the spreading process. In Fig. \ref{fig:phase_space}, we show that for a given network structure and constant $\eta=0.12$, higher values of $\theta$ correspond to a lower critical mobility rate and a larger invasion regime phase space.
On the other side, by assuming a negative value of $\theta$, therefore a more homogeneous distribution of the mobility flows across the network, the global spreading regime is suppressed. 
In both cases, it is remarkable that the numerical simulations show a very good agreement with the theoretical value of the threshold (black solid line in Fig.\ref{fig:phase_space}). Also in this case we considered uncorrelated scale-free networks with $V=10^5$ nodes, and exponent $\gamma=-2.1$. Each point is averaged in $2 \times 10^3$ Monte Carlo simulations. In the Supplemental Material, we report the full details of the numerical simulations methods.

In conclusion, prompted by empirical findings, we derived a general framework to study spreading processes in metapopulation systems where the individual contact rates scale with subpopulation sizes.  The effects of local properties of the subpopulations in RD processes, including different local mixing patterns, have been studied in previous works \cite{Shen2012, Wang2013, Lund2013, Tanaka2014, Gong2014, Mata2014} but they were generally limited to simplified assumptions on the local contact structure, such as considering only two different contact rates \cite{Lund2013, Wang2013, Tanaka2014}, and by always assuming a constant diffusion rate \cite{Shen2012, Wang2013, Gong2014}.
Some recent papers have also considered a power law distribution of the infectious rates in a metapopulation model \cite{Takaguchi2014, Gong2014}. However, a comprehensive framework that takes into account the interplay between the heterogeneities of both mobility flows and contact rates was still missing.
We have shown that the heterogeneity of the contact rates, introduced by the scaling behavior, promotes the epidemic spreading. Furthermore, we have shown that such effect is enhanced when the distribution of the mobility flows between subpopulations is heterogeneous, as observed in real mobility networks. 
Our results represent the first step towards a better analytical understanding of contagion processes, as the spreading of infectious diseases and information, in structured subpopulations.

\begin{acknowledgments}
This work has been partially funded by the EC FET-Proactive Project MULTIPLEX (Grant No. 317532) to MT. We thank A. Vespignani for helpful discussions, insights, and comments, and N. Samay for help in drafting the figures.
\end{acknowledgments}

%

\end{document}